\documentclass[5p,twocolumn,authoryear,times]{elsarticle}

\usepackage{lineno,hyperref}
\modulolinenumbers[5]
\hypersetup{colorlinks,citecolor=blue,linkcolor=blue,urlcolor=blue}

\journal{Astronomy \& Computing}

\usepackage{graphicx}	
\usepackage{amsmath}	
\usepackage{amssymb}	
\usepackage{bm}

\usepackage{natbib} 
\usepackage[usenames,dvipsnames]{color}
\usepackage{verbatim}
\usepackage{multirow}
\usepackage{booktabs}
\usepackage{subfigure}
\usepackage{color, colortbl}

\usepackage[T1]{fontenc}
\usepackage[utf8]{inputenc}
\usepackage[tracking=smallcaps]{microtype}
\SetTracking[no ligatures = f]{encoding = *,shape = sc}{0}

\usepackage{listings}
\usepackage[normalem]{ulem}
\usepackage[font=footnotesize,labelfont=bf]{caption}
\usepackage{footnote}
\makesavenoteenv{tabular}
\makesavenoteenv{table}
\usepackage{supertabular}
\usepackage{threeparttable}
\usepackage{longtable}
\usepackage{rotating}
\usepackage{lscape}

\usepackage{lineno}

\usepackage{bold-extra}
\usepackage{slantsc}
\usepackage{times}

\usepackage{blindtext}
\makeatletter
\newcommand{\removelatexerror}{\let\@latex@error\@gobble}
\makeatother

\usepackage[]{algorithm2e}
\usepackage{siunitx}
\usepackage{dsfont}
\usepackage{bbm}

\usepackage{tikz,pgfplots}

\let\oldFootnote\footnote
\newcommand\nextToken\relax

\renewcommand\footnote[1]{%
    \oldFootnote{#1}\futurelet\nextToken\isFootnote}

\newcommand\isFootnote{%
    \ifx\footnote\nextToken\textsuperscript{,}\fi}
\usepackage[multiple]{footmisc}

\definecolor{babypink}{rgb}{0.96, 0.76, 0.76}
\definecolor{beaublue}{rgb}{0.74, 0.83, 0.9}
\definecolor{mediumturquoise}{rgb}{0.28, 0.82, 0.8}
\definecolor{mossgreen}{rgb}{0.68, 0.87, 0.68}
\definecolor{mustard}{rgb}{1.0, 0.86, 0.35}
\definecolor{olivine}{rgb}{0.6, 0.73, 0.45}
\definecolor{orchid}{rgb}{0.85, 0.44, 0.84}
\definecolor{palecerulean}{rgb}{0.61, 0.77, 0.89}
\definecolor{palegold}{rgb}{0.9, 0.75, 0.54}
\definecolor{paleplum}{rgb}{0.8, 0.6, 0.8}
\definecolor{palespringbud}{rgb}{0.93, 0.92, 0.74}
\definecolor{pastelgray}{rgb}{0.81, 0.81, 0.77}
\definecolor{pastelviolet}{rgb}{0.8, 0.6, 0.79}
\definecolor{pastelred}{rgb}{1.0, 0.41, 0.38}
\definecolor{pearl}{rgb}{0.94, 0.92, 0.84}
\definecolor{pistachio}{rgb}{0.58, 0.77, 0.45}	
\definecolor{teal}{rgb}{0.0, 0.5, 0.5}
\definecolor{tiffanyblue}{rgb}{0.04, 0.73, 0.71}
\definecolor{turquoise}{rgb}{0.19, 0.84, 0.78}
\definecolor{verdigris}{rgb}{0.26, 0.7, 0.68}
\definecolor{lightgray}{gray}{0.9}

\definecolor{blue_mlp}{rgb}{0.122, 0.467, 0.706}
\definecolor{orange_mlp}{rgb}{0.957, 0.643, 0.376}
\definecolor{red_mlp}{rgb}{0.804, 0.361, 0.361}

 {\pagebreak[4]\global\pdfpageattr\expandafter{\the\pdfpageattr/Rotate 90}}%
 {\pagebreak[4]\global\pdfpageattr\expandafter{\the\pdfpageattr/Rotate 0}}%
 
\usepackage{listings}
\lstnewenvironment{codebkg}{%
  \lstset{
    escapechar=\&,
    basicstyle=\ttfamily\scriptsize,
    backgroundcolor=\color{lightgray},
    frame=single,
    breaklines=true,    
    framerule=0pt,
    columns=fullflexible
  }
}
{}

\begin{document}
\begin{frontmatter}

\title{Solar flare forecasting with foundational transformer models \\across image, video, and time-series modalities}

\author[1]{S. Riggi\corref{cor}}%
\ead{simone.riggi@inaf.it}
\author[1]{P. Romano}
\author[2]{A. Pilzer}
\author[1]{U. Becciani}

\cortext[cor]{Corresponding author}%
\address[1]{INAF - Osservatorio Astrofisico di Catania, Via Santa Sofia 78, 95123 Catania, Italy}%
\address[2]{NVIDIA AI Technology Center, Italy}%

\begin{abstract}
We present a comparative study of transformer-based architectures for solar flare forecasting using heterogeneous data modalities, including images, video sequences, and time-series observations. Our analysis evaluates three recent foundational models $-$ \textit{SigLIP2} for image encoding, \textit{VideoMAE} for spatio-temporal video representation, and \textit{Moirai2} for multivariate time-series forecasting $-$ applied to publicly available datasets of solar magnetograms from the SDO/HMI mission and soft X-ray fluxes acquired by GOES satellites.
All models are trained and validated under consistent data splits and evaluation criteria, with the goal of assessing the strengths and limitations of transformer backbones across spatial and temporal representations of solar activity. We investigate multiple loss formulations (weighted BCE, focal, and score-oriented) and training balance strategies to mitigate class imbalance typical of flare datasets.
Results show that while both SigLIP2 and VideoMAE achieve typical performance on image and video data (True Skill Statistic TSS$\sim$0.60-0.65), the time-series model Moirai2 reaches superior forecasting skill (TSS$\sim$0.74) using irradiance-based temporal evolution alone. These findings highlight the potential of pretrained transformer architectures and cross-modal learning for advancing operational space weather forecasting, paving the way toward unified multimodal models that integrate visual and temporal information.
\end{abstract}


\end{frontmatter}


\section{Introduction}
\label{sec:intro}
Solar flares are among the most energetic manifestations of magnetic energy release in the solar atmosphere. They are powered by magnetic reconnection — the rapid topological reconfiguration of coronal magnetic field lines — which converts magnetic free energy accumulated in active regions into plasma heating, particle acceleration, and broadband electromagnetic emission, from hard X-rays to radio wavelengths \citep{Shi11, Ben17}. The reconnection process typically occurs in the corona above magnetic polarity inversion lines where strong shear and electric currents develop \citep{Pri02, Tor19}. The released energy propagates along magnetic loops toward the lower atmosphere, producing impulsive chromospheric brightenings and heating of dense plasma that fills coronal loops through chromospheric evaporation.

This magnetic energy build-up is driven by the continuous emergence, twisting, and shearing of photospheric magnetic flux. Observations and nonlinear force-free field extrapolations have shown that flare-productive active regions are characterized by high magnetic complexity and helicity, strong shear along polarity inversion lines, and the presence of magnetic flux ropes or $\delta$-sunspots \citep[e.g.,][]{Rom14, Rom18, Rom24}. These magnetic configurations, identifiable through their footprints in photospheric magnetograms, favour the storage of large amounts of free magnetic energy, which can be rapidly released through reconnection once critical thresholds of current density or magnetic stress are exceeded. The onset of a flare is often accompanied by coronal restructuring and, in many cases, by the ejection of magnetized plasma in the form of a coronal mass ejection (CME), whose propagation through the heliosphere can drive interplanetary shocks and strong space-weather disturbances \citep{Zha01, Tem21}.

The impact of these eruptive phenomena on Earth and near-Earth space can affect both technological systems and human health. Major consequences include disruption of radio communications, damage to satellites and spacecraft, increased radiation exposure for astronauts and high-altitude flights, geomagnetically induced currents (GICs) in power grids, and disturbances in navigation and communication systems \citep{Pul17, Kil17}. 

Solar flares are commonly categorized by thresholds of the soft X-ray maximum peak flux (MPF) in the 0.1–0.8 nm band, where M-class (10$^{-5}$ < MPF < 10$^{-4}$ W m$^{-2}$) and X-class (MPF > 10$^{-4}$ W m$^{-2}$) flares indicate stronger events that are often, but not always, associated with the most geoeffective CMEs and particle storms. Accurate forecasting of solar flares and CMEs is therefore a key objective in space-weather research, enabling mitigation of their potentially adverse effects. 

Over the past decades, flare forecasting has evolved through several methodological stages, progressing from empirical and statistical approaches to physics-based modeling and, more recently, to machine learning techniques. The advent of high–spatial and high–temporal resolution multi-wavelength observations from modern solar observatories $-$ such as NASA's Solar Dynamics Observatory (SDO, \citealt{SDO}), ESA/NASA's Solar Orbiter \citep{Muller2020}, NASA's Parker Solar Probe (PSP, \citealt{ParkesSolarProbe}), ESA's Proba-3 \citep{Zhukov2025}, the Advanced Space-based Solar Observatory (ASO-S, \citealt{Gan2023}), ISRO's Aditya-L1 \citep{Tripathi2022}, and NOAA's Geostationary Operational Environmental Satellites (GOES) $-$ has enabled the development of deep learning models capable of forecasting directly from images, sequences, videos, and multimodal data. These approaches minimize the need for manual feature engineering and can potentially uncover complex spatio-temporal patterns associated with solar activity.

In this context, the goal of this work is to explore how recent state-of-the-art transformer-based foundational architectures, designed for different data modalities (images, videos, time series), perform on the solar flare forecasting problem. To support future research and reproducibility within the community, we carried out our analysis on a publicly available dataset and released both the developed software and fine-tuned models as open resources \footnote{\url{https://github.com/inaf-oact-ai/solar-flare-forecaster/}}\footnote{\url{https://huggingface.co/inaf-oact-ai}}. 
While transformer-based video models have only recently been applied to flare forecasting \citep{Li2025}, this study represents, to the best of our knowledge, one of the first explorations of foundational models for time-series in this domain. 

The paper is organized as follows. Section~\ref{sec:literature-review} reviews related work, with emphasis on recent advances in deep learning architectures. Section~\ref{sec:dataset} describes the datasets used for all data modalities and the procedures adopted for generating the training, validation, and test splits. Section~\ref{sec:method} provides an overview of the foundational models considered and describes how they were adapted to the solar flare forecasting task, along with details of the training and evaluation strategies. The obtained results are presented and discussed in Section~\ref{sec:results}, and the conclusions and future directions are summarized in Section~\ref{sec:summary}.

\section{Related work}
\label{sec:literature-review}
Traditional solar flare forecasting has relied on handcrafted physical and empirical predictors, such as magnetic field features, historical flare activity, and statistical models. The recent availability of high-resolution solar imagery and magnetogram data (e.g., from SDO/HMI\footnote{Helioseismic and Magnetic Imager}) has facilitated the adoption of deep learning techniques for this task. In this section, we review the main deep learning approaches proposed to date, organized by data modality, and summarize their forecasting performance in terms of the \emph{True Skill Statistic} (TSS) and the \emph{Heidke Skill Score} (HSS) metrics (see Section~\ref{subsec:model-training} and~\ref{appendix:metrics}).

\subsection{Image-based approaches}
Image-based methods have long represented the most direct approach to solar flare forecasting, exploiting the rich spatial information encoded in photospheric and coronal images. Most studies rely on SDO/HMI line-of-sight (LOS) or vector magnetograms, provided either as full-disk maps or as cropped active region (AR) patches. The general goal is to determine whether an AR will produce a flare above a given class threshold (e.g., $\ge$C, M, or X) within a specified prediction window, typically 24 hours.

Early work in this domain established the foundation for deep-learning-based flare prediction through convolutional neural networks (CNNs), which demonstrated the capacity to extract meaningful magnetic field features from magnetograms. For instance, \citet{Huang2018} designed a custom CNN trained on LOS magnetograms from SOHO/MDI and SDO/HMI to investigate flare predictability across multiple thresholds and forecast horizons, achieving a TSS of 0.66 and an HSS of 0.14 for M-class and above (M+) forecasting over a 24-hour horizon. Building upon this, \citet{Zheng2019} introduced a hybrid architecture of three binary CNN classifiers, inspired by VGGNet, for multi-class flare prediction. Using SDO/HMI magnetograms, they reported mean TSS scores ranging from 0.53 to 0.55 across flare classes, highlighting the discriminative potential of CNN features for different flare intensity levels.

Subsequent studies began addressing key challenges such as class imbalance and solar-cycle variability. \citet{Deng2021} incorporated a Generative Adversarial Network (GAN) to generate synthetic magnetograms, thereby augmenting the training data and reducing imbalance between flare and non-flare samples. Their hybrid CNN, trained separately for the rising and declining phases of the solar cycle, achieved mean TSS values of 0.65 (C), 0.65 (M), and 0.76 (X), marking a notable improvement over earlier CNN-based methods.

With the success of the attention mechanism in computer vision \citep{Vaswani2017}, subsequent models integrated spatial attention blocks or adopted Vision Transformer (ViT) \citep{Dosovitskiy2020} backbones to enhance feature extraction. \citet{Pandey2023} introduced an attention-augmented CNN trained directly on full-disk SDO/HMI magnetograms for $\ge$M1.0-class flare prediction, achieving average scores of TSS = 0.54 $\pm$ 0.03 and HSS = 0.37 $\pm$ 0.07 $-$ significantly outperforming standard CNNs without attention. In a systematic evaluation, \citet{Yan2024} compared conventional CNNs, attention-enhanced variants (SE, CBAM, and ECA blocks), and a pre-trained ViT Base/16 architecture. Their analysis showed that attention modules did not lead to statistically significant gains over baseline CNNs across all metrics, although the CNN-SE variant yielded the best overall results, with TSS = 0.68 $\pm$ 0.13 and HSS = 0.69 $\pm$ 0.13 for C+ forecasting, and TSS = 0.56 $\pm$ 0.10 and HSS = 0.32 $\pm$ 0.07 for M+ forecasting.

The exploration of image modalities beyond magnetograms has also expanded the scope of flare forecasting. \cite{Sun2023} applied CNN-based architectures (ResNet18, AlexNet, SqueezeNet) to SDO/AIA extreme ultraviolet (EUV) images at six wavelengths (94, 131, 171, 193, 211, and 335~\r{A}), finding that the 94~\r{A} channel provided the strongest single-wavelength performance (TSS = 0.92), while multi-wavelength fusion achieved TSS $\sim$ 0.75. More recently, \cite{Francisco2025} proposed a patch-distributed CNN (P-CNN, based on EfficientNetV2-S) to forecast C+ and M+ flares from full-disk SDO/HMI and SDO/AIA images. Their EUV-based models achieved slightly higher scores (TSS = 0.61, HSS = 0.42) than the magnetogram-based ones, while the regional patch configuration (TSS = 0.43, HSS = 0.22) demonstrated potential for localizing flare-productive regions without explicit AR labeling.

Other studies, such as \citet{Legnaro2025}, explored the classification of sunspot magnetic type in solar active-region cutouts using ResNet18 and DeiT (Data-Efficient Image Transformer) architectures, demonstrating that combining magnetogram and continuum images can significantly enhance model performance.

Together, these studies outline a steady evolution of image-based flare forecasting $-$ from early CNN models to hybrid and transformer-augmented architectures $-$ and reveal both the promise and the limits of spatial-only approaches. Despite meaningful progress, performance remains sensitive to dataset balance, magnetic complexity representation, and the inclusion (or lack) of temporal and multi-wavelength information, motivating the shift toward video-based and multimodal forecasting strategies.

\subsection{Spatio-temporal and video-based forecasting approaches}
Building upon the advances of image-based flare forecasting, recent research has shifted toward modeling the AR temporal evolution through spatio-temporal or video-based approaches. These methods aim to capture how magnetic structures evolve and accumulate energy prior to flare onset, integrating both spatial morphology and temporal dynamics. By combining convolutional and sequential processing $-$ through CNN–recurrent network hybrids, 3D convolutional networks, or transformer architectures $-$ they attempt to represent the buildup of magnetic complexity that static models cannot fully resolve.

An early demonstration of this paradigm was provided by \citet{Guastavino2022}, who applied a Long-term Recurrent Convolutional Network (LRCN) to 24-hour SDO/HMI magnetogram sequences (40 frames at a 36-minute cadence). The model combined CNN-based spatial encoders with an Long short-term memory (LSTM) module to learn temporal dependencies, achieving mean TSS scores of 0.55~$\pm$~0.05 and 0.68~$\pm$~0.09 for C+ and M+ forecasting, respectively. 

\cite{Sun2022} employed 3D CNNs to model 24-hour SHARP magnetogram sequences (15 frames at a 96-minute cadence) for C+ and M+ flare prediction within the next 24 hours. To address class imbalance, they adopted a focal loss function and systematically explored the impact of the focal parameter $\gamma$. Their best configuration, with $\gamma = 2$, achieved TSS = 0.826 and HSS = 0.667 for M+ forecasting.

Subsequent efforts explored architectures capable of capturing more complex temporal correlations and multi-wavelength context. \citet{Francisco2024} proposed VideoLENS, an architecture integrating a C3D block with multiple 3D convolutional layers and a Local-TimeSeries module combining multi-head attention and LSTM layers. Trained on 24-hour sequences (13 frames at a 2-hour cadence) of full-disk SDO/AIA EUV images at 193, 211, and 94~\r{A}, VideoLENS achieved TSS = 0.65~$\pm$~0.01 and HSS = 0.50~$\pm$~0.01 for M+ forecasting $-$ an improvement of about 5\% relative to single-channel models. Their results established that incorporating temporal evolution, even over relatively short windows, can enhance predictive performance compared to static image classifiers.

Further refining temporal modeling, \citet{Xu2025} developed a hybrid CNN–TCN framework that combined spatial encoding with Temporal Convolutional Networks (TCN) to model the progressive evolution of SDO/HMI magnetograms, supplemented by magnetic field parameters. Their model achieved TSS = 0.85~$\pm$~0.07 and HSS = 0.75~$\pm$~0.10 for 24-hour C+ and M+ predictions $-$ among the highest scores reported to date $-$ demonstrating the effectiveness of TCNs in representing medium-term dependencies in solar magnetic activity.

The most recent advances introduced transformer-based video models capable of learning long-range spatio-temporal dependencies. \citet{Li2025} employed a Multiscale Video Transformer (MViT) trained on 24-hour magnetogram sequences (40 frames at 36-minute cadence) from SDO/SHARP, SDO/HMI, and FMG/ASO-S instruments. Their analysis compared single- and multi-AR samples, showing that transformer models consistently outperformed CNN baselines. Performance was notably higher for single-AR sequences (TSS = 0.74~$\pm$~0.06, HSS = 0.61~$\pm$~0.12) compared with mixed ones (TSS = 0.69~$\pm$~0.05, HSS = 0.58~$\pm$~0.08), underscoring the benefits of isolating individual ARs during training.

\cite{Grim2024} further extended transformer-based approaches by applying Improved Multiscale Vision Transformers (MViTv2) to 24-hour SDO/HMI magnetogram sequences (16 frames at a 96-minute cadence) for 24- and 48-hour flare forecasting. Using a weighted loss and oversampling to mitigate class imbalance, their best M+ model achieved TSS = 0.70~$\pm$~0.04 and HSS = 0.25~$\pm$~0.01 for the 24-hour horizon.

Collectively, these studies mark a clear progression from static image classifiers to architectures capable of capturing the dynamic evolution of solar magnetic fields. While performance gains have been moderate compared to single-image models, spatio-temporal approaches provide a more physically grounded framework for flare forecasting and set the stage for multimodal models that jointly exploit temporal, spectral, and contextual information from multiple instruments.

\begin{figure*}[!htb]
\centering%
\subtable[No-Flare \textit{(AR=2563, t=20160714 00:00 UT)}]{\includegraphics[scale=0.63]{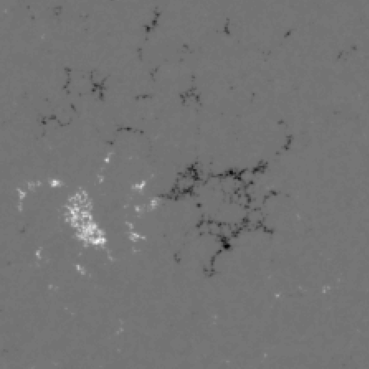}}
\subtable[C-class \textit{(AR=2396, t=20120806 19:00 UT)}]{\includegraphics[scale=0.63]{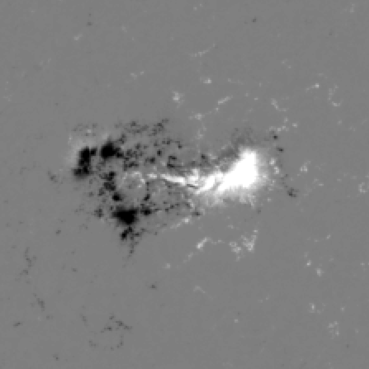}}
\subtable[X-class \textit{(AR=2673, t=20170905 06:00 UT)}]{\includegraphics[scale=0.63]{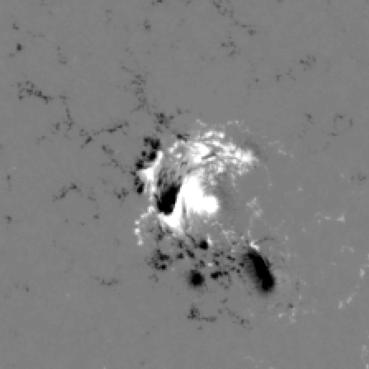}}%
\caption{Sample of HMI magnetograms from the image dataset, illustrating examples of the NONE (left), C-class (middle), and X-class (right) categories. The NONE example corresponds to an AR with no flares stronger than the C class occurring within the following 24 hours. For each image, the associated AR identifier and observation times are reported below.}%
\label{fig:hmi-sample-images}
\end{figure*}

\subsection{Time series-based approaches}
Parallel to the development of image and video models, another major line of research has focused on forecasting solar flares directly from time series of magnetic and radiative parameters that characterize ARs. Rather than modeling spatial morphology, these approaches capture the temporal evolution of key physical quantities, such as those derived from the SHARP or HMI datasets, to identify dynamical precursors of flare activity. Deep learning methods including recurrent networks (RNNs, LSTMs, GRUs), one-dimensional convolutional networks (1D-CNNs), and more recently, transformer-based architectures have been employed to uncover such temporal dependencies and correlations.

Early contributions demonstrated the promise of recurrent architectures for this task. \cite{Chen2019} trained a two-layer LSTM model on multivariate time series of 20 SHARP parameters sampled at 1-hour cadence, predicting flares across multiple thresholds and forecast horizons ranging from 1 to 72 hours. Their model maintained stable performance for short-term forecasts but exhibited a gradual decline for longer lead times, achieving TSS = 0.68 and HSS = 0.68 for 24-hour M+ forecasting. This work highlighted both the effectiveness of LSTMs in modeling flare-related temporal dynamics and the inherent difficulty of long-horizon prediction.

Building on this foundation, \citet{Abduallah2023} introduced \emph{SolarFlareNet}, a more sophisticated hybrid model that integrates 1D-CNN, LSTM, and transformer encoder blocks to capture both short- and long-term dependencies in the SHARP parameters. Trained on 108-minute sequences of nine magnetic parameters sampled at 12-minute cadence, the model forecasted flares at various intensity thresholds ($\ge$C, $\ge$M, $\ge$M5.0) and horizons (24, 48, and 72 hours). Using the start of each sequence as the reference time for labeling,\footnote{This differs from our approach, in which the last time step defines the forecast reference.} their best model achieved a mean TSS of 0.84 $\pm$ 0.03 for 24-hour M+ forecasting $-$ one of the highest values reported for time-series methods to date.

More recently, \citet{Donahue2024} applied a pure transformer architecture to 24-hour and 48-hour sequences of 18 SHARP parameters, directly leveraging the transformer’s ability to model long-range temporal dependencies without recurrent connections. For a 24-hour horizon, their M+ forecasting model reached TSS = 0.66 and HSS = 0.16, confirming the growing potential of transformer-based encoders for time-domain flare prediction.

For completeness, we also note several recent studies that have focused on high-cadence time series of soft X-ray fluxes (XRS) recorded by the GOES. Unlike the previously discussed works, which frame flare prediction as a classification problem (i.e., predicting the flare class), these analyses approach the task as a multi-step regression problem, aiming to forecast the future evolution of the XRS flux itself or related observables over specific time horizons. 

For instance, \citet{Dei2020} compared three neural architectures $-$ 1D CNN, LSTM, and N-BEATS $-$ for forecasting X-ray fluxes at lead times ranging from 1 to 5 hours, using historical windows of 1, 2, and 3 times the prediction interval. Among these, the N-BEATS model consistently achieved the lowest mean squared error (MSE$\sim$0.001 for the largest prediction window), outperforming both baseline estimators and CNN/LSTM networks. 

More recently, \citet{vanDerSande2025} evaluated a set of traditional machine learning and deep learning models (Linear, MLP, Random Forest, and CNN) to predict XRS fluxes at horizons up to $\sim$48~h. Their models used a diverse set of input sources, including magnetograms, EUV images at multiple wavelengths, and scalar magnetic features. Forecasting accuracy was found to improve with longer prediction windows, with their best CNN model reaching MSE~$\sim$~0.3 after 24~h. From the predicted flux values, they also derived binary classification metrics for $\ge$M1-class events, obtaining TSS and HSS values of approximately 0.6 for forecasting windows beyond 24~h.

\begin{table*}[!htb]
\centering\setlength{\aboverulesep}{0pt}\setlength{\belowrulesep}{0pt}
\footnotesize%
\setlength{\extrarowheight}{2pt}
\caption{Number of samples per solar flare class present in each dataset used in this work.}
\begin{tabular}{llccccccccccccc}
\hline%
\hline%
\rowcolor{palegold}
\multirow{2}{*}{\cellcolor{palegold}}%
& & & 
\multicolumn{2}{c}{No-Flare} & \multicolumn{2}{c}{C-class} & \multicolumn{2}{c}{M-class} & \multicolumn{2}{c}{X-class} & \multicolumn{2}{c}{NONE} & \multicolumn{2}{c}{M+}\\
\arrayrulecolor{palegold}\midrule\noalign{\vspace*{-\cmidrulewidth}}%
\arrayrulecolor{black}%
\cmidrule(lr){4-5}%
\cmidrule(lr){6-7}%
\cmidrule(lr){8-9}%
\cmidrule(lr){10-11}%
\cmidrule(lr){12-13}%
\cmidrule(lr){14-15}%
\rowcolor{palegold}
\multirow{-2}{*}{Data} &  \multirow{-2}{*}{Sample} & \multirow{-2}{*}{All} &  $N$ & $f$(\%) & $N$ & $f$(\%) & $N$ & $f$(\%) & $N$ & $f$(\%) & $N$ & $f$(\%) & $N$ & $f$(\%)\\
\midrule
HMI & all & 950,047 & 759,465 & 79.9 & 161,059 & 17.0 & 26,680 & 2.8 & 2,843 & 0.3 & 920,524 & 96.9 & 29,523 & 3.1\\
\cline{2-15}
images & train & 759,357 & 610,108 & 80.3 & 125,854 & 16.6 & 21,141 & 2.8 & 2,254 & 0.3 & 735,962 & 96.9 & 23,395 & 3.1\\
 & train\_ds & 299,249 & 150,000 & 50.1 & 125,854 & 42.1 & 21,141 & 7.1 & 2,254 & 0.8 & 275,854 & 92.2 & 23,395 & 7.8\\
 & train\_ds\_bal & 48,395 & 12,500 & 25.8 & 12,500 & 25.8 & 21,141 & 43.7 & 2,254 & 4.7 & 25,000 & 51.7 & 23,395 & 48.3\\
 & cv & 95,933 & 76,142 & 79.4 & 16,975 & 17.7 & 2,578 & 2.7 & 238 & 0.2 & 93,117 & 97.1 & 2,816 & 2.9\\
 & cv\_ds\_bal & 5,816 & 1,500 & 25.8 & 1,500 & 25.8 & 2,578 & 44.3 & 238 & 4.1 & 3,000 & 51.6 & 2,816 & XXX\\
 & test & 94,757 & 73,215 & 77.2 & 18,230 & 19.2 & 2,961 & 3.1 & 351 & 0.4 & 91,445 & 96.5 & 3,312 & 3.5\\
\hline%
HMI & all & 631,366 & 503,293 & 79.7 & 109,062 & 17.3 & 17,212 & 2.7 & 1,799 & 0.3 & 612,355 & 97.0 & 19,011 & 3.0\\
\cline{2-15}
videos & train & 504,526 & 404,790 & 80.2 & 84,365 & 16.7 & 13,979 & 2.8 & 1,392 & 0.3 & 489,155 & 97.0 & 15,371 & 3.0\\
($\Delta t$=72 min) & train\_ds & 199,736 & 100,000 & 50.1 & 84,365 & 42.2 & 13,979 & 7.0 & 1,392 & 0.7 & 184,365 & 92.3 & 15,371 & 7.7\\
 & train\_ds\_bal & 30,771 & 7,700 & 25.0 & 7,700 & 25.0 & 13,979 & 45.4 & 1,392 & 4.5 & 15,400 & 50.0 & 15,371 & 50.0\\
 & cv & 62,730 & 48,714 & 77.7 & 12,094 & 19.3 & 1,773 & 2.8 & 149 & 0.2 & 60,808 & 96.9 & 1,922 & 3.1\\
 & cv\_ds\_bal & 3,922 & 1,000 & 25.5 & 1,000 & 25.5 & 1,773 & 45.2 & 149 & 3.8 & 2,000 & 51.0 & 1,922 & 49.0\\
& test & 64,110 & 49,789 & 77.7 & 12,603 & 19.7 & 1,460 & 2.3 & 258 & 0.4 & 62,392 & 97.3 & 1,718 & 2.7\\
\hline%
HMI & all & 774,687 & 617,791 & 79.7 & 133,028 & 17.2 & 21,597 & 2.8 & 2,271 & 0.3 & 750,819 & 96.9 & 23,868 & 3.1\\%
\cline{2-15}
videos & train & 618,864 & 496,211 & 80.2 & 103,501 & 16.7 & 17,380 & 2.8 & 1,772 & 0.3 & 599,712 & 96.9 & 19,152 & 3.1\\%
($\Delta t$=36 min) & train\_ds & 222,653 & 100,000 & 44.9 & 103,501 & 46.5 & 17,380 & 7.8 & 1,772 & 0.8 & 203,501 & 91.4 & 19,152 & 8.6\\
& train\_ds\_bal & 38,352 & 9,600 & 25.0 & 9,600 & 25.5 & 17,380 & 45.3 & 1,772 & 4.6 & 19,200 & 50.1 & 19,152 & 49.9\\
& cv & 77,924 & 61,122 & 78.4 & 14,450 & 18.5 & 2,159 & 2.8 & 193 & 0.2 & 75,572 & 97.0 & 2,352 & 3.0\\
& cv\_ds\_bal & 4,752 & 1,200 & 25.3 & 1,200 & 25.3 & 2,159 & 45.4 & 193 & 4.1 & 2,400 & 50.5 & 2,352 & 49.5\\
& test & 77,899 & 60,458 & 77.6 & 15,077 & 19.4 & 2,058 & 2.6 & 306 & 0.4 & 75,535 & 97.0 & 2,364 & 3.0\\
\hline%
XRS & all & 20,755 & 10,307 & 49.7 & 6,964 & 33.6 & 3,081 & 14,8 & 403 & 1.9 & 17,271 & 83.2 & 3,484 & 16.8\\
\cline{2-15}
series  & train & 14,529  & 7,215 & 49.7 & 4,875 & 33.6 & 2,157 & 14.8 & 282 & 1.9 & 12,090 & 83.2 & 2,439 & 16.8\\
($T$=24 h) & cv & 2,076 & 1,031 & 49.7 & 697 & 33.6 & 308 & 14.8 & 40 & 1.9  & 1,728 & 83.2 & 348 & 16.8\\
& test & 4,150 & 2,061 & 49.7 & 1,392 & 33.6 & 616 & 14.8 & 81 & 1.9  & 3,453 & 83.2 & 697 & 16.8\\
\hline%
XRS & all & 41,517 & 23,583 & 56.8 & 13,283 & 32.0 & 4,207 & 10.1 & 444 & 1.1 & 36,866 & 88.8 & 4,651 & 11.2\\%
\cline{2-15}
series & train & 29,062 & 16,508 & 56.8 & 9,298 & 32.0 & 2,945 & 10.1 & 311 & 1.1 & 25,806 & 88.8  & 3,256 & 11.2\\%
($T$=12 h) & cv & 4,151 & 2,358 & 56.8 & 1,328 & 32.0 & 421 & 10.1 & 44 & 1.1 & 3,686 & 88.8 & 465 & 11.2\\
& test & 8,303 & 4,716 & 56.8  & 2,657 & 32.0 & 841 & 10.1 & 89 & 1.1 & 7,373 & 88.8 & 930 & 11.2\\
\hline%
\hline%
\end{tabular}
\label{tab:dataset}
\end{table*}

\section{Dataset}
\label{sec:dataset}
In this work, we employ three distinct data modalities as inputs to the flare forecasting models: images, videos, and time series. Table~\ref{tab:dataset} summarizes the number of available samples and subsets for each modality. The following sections provide detailed descriptions of the original observations and the data preparation workflow, including event annotation procedures and the generation of training, validation, and test splits.

\subsection{Image dataset}
\label{subsec:image-data}
The image dataset used in this work was compiled by \cite{Boucheron2023} from full-disk LOS solar photospheric magnetograms acquired by the Helioseismic and Magnetic Imager (HMI) aboard NASA's SDO between 1 May 2010 and 31 December 2018, with a cadence of 720 s. The original full-disk HMI observations were cropped around the heliographic position (within $\pm60^{\circ}$ in latitude and longitude) of 1,570 ARs reported by the National Oceanic and Atmospheric Administration (NOAA), yielding 600$\times$600-pixel cutouts (approximately 300"$\times$300"), large enough to encompass the most extended ARs. The final dataset comprises 950,047 magnetogram images, each centered on a NOAA-identified AR.

For this study, we adopted the lower-resolution version of the dataset, where the full-resolution images were resized to 224$\times$224 pixels and preprocessed following the pipeline described in \cite{Boucheron2023}:
\begin{itemize}
\item offset pixel intensities $I$ by 2550, i.e., $I \rightarrow I + 2550$;
\item clip intensities to the range [0, 5100];
\item normalize values to [0, 255];
\item downsample intensities to 8-bit depth (uint8).
\end{itemize}
Image labeling was performed using the Space Weather Prediction Center (SWPC) Event Reports, which list solar flares detected by the GOES and their associated NOAA ARs. Each HMI magnetogram was assigned a categorical label indicating whether a flare of class $\ge$C1.0 occurred within the subsequent 24 hours relative to the image timestamp. Specifically, images were labeled as No-Flare if no flares above the C class occurred, or otherwise by the major flare class (C, M, or X), disregarding subtypes. Representative examples of HMI images labeled as No-Flare, C, and X are shown in Fig.~\ref{fig:hmi-sample-images}.
 
The total number of images per flare class is reported in Table~\ref{tab:dataset} (first row). Following \cite{Boucheron2023}, the dataset includes predefined training, validation (cv), and test subsets, whose class distributions are listed in rows 2, 5, and 7 of the same table. The dataset is markedly imbalanced, dominated by No-Flare samples (~80\%), followed by C-class (17–19\%), M-class (~3\%), and the rare X-class (~0.3\%). We additionally report aggregated class-grouped versions labeled as NONE (combining No-Flare and C-class samples) and M+ (combining M- and X-class samples).

To mitigate this imbalance, we created additional downsampled datasets by randomly reducing the number of No-Flare and C-class images while retaining at least one image per AR (so ARs with more images were more aggressively downsampled). The target number of images for each downsampled class was set approximately equal to the combined number of M+ samples. The resulting datasets, denoted with a "\_ds" suffix (e.g., \textit{train\_ds}), and their class abundances are reported in Table~\ref{tab:dataset}. More strongly balanced versions, labeled "\_ds\_bal" (e.g., \textit{train\_ds\_bal}), were produced by further downsampling No-Flare and C-class data to achieve a nearly 1:1 ratio between positive and negative examples. Unless otherwise specified, only the downsampled datasets were used for training the image-based forecasting models.
 
Finally, to assess statistical variability, we generated five-fold (K=5) cross-validation splits from the original dataset by randomly reassigning all images from each AR to a single split (ensuring that no AR appears in multiple folds). The overall class distributions were preserved to within $\sim$1\% of the original dataset on average, with a standard deviation across folds below 2\%. These K-fold splits were used to estimate the uncertainty of forecasting metrics reported for the image-based models.

\begin{figure*}[!htb]
\centering%
\includegraphics[scale=0.5]{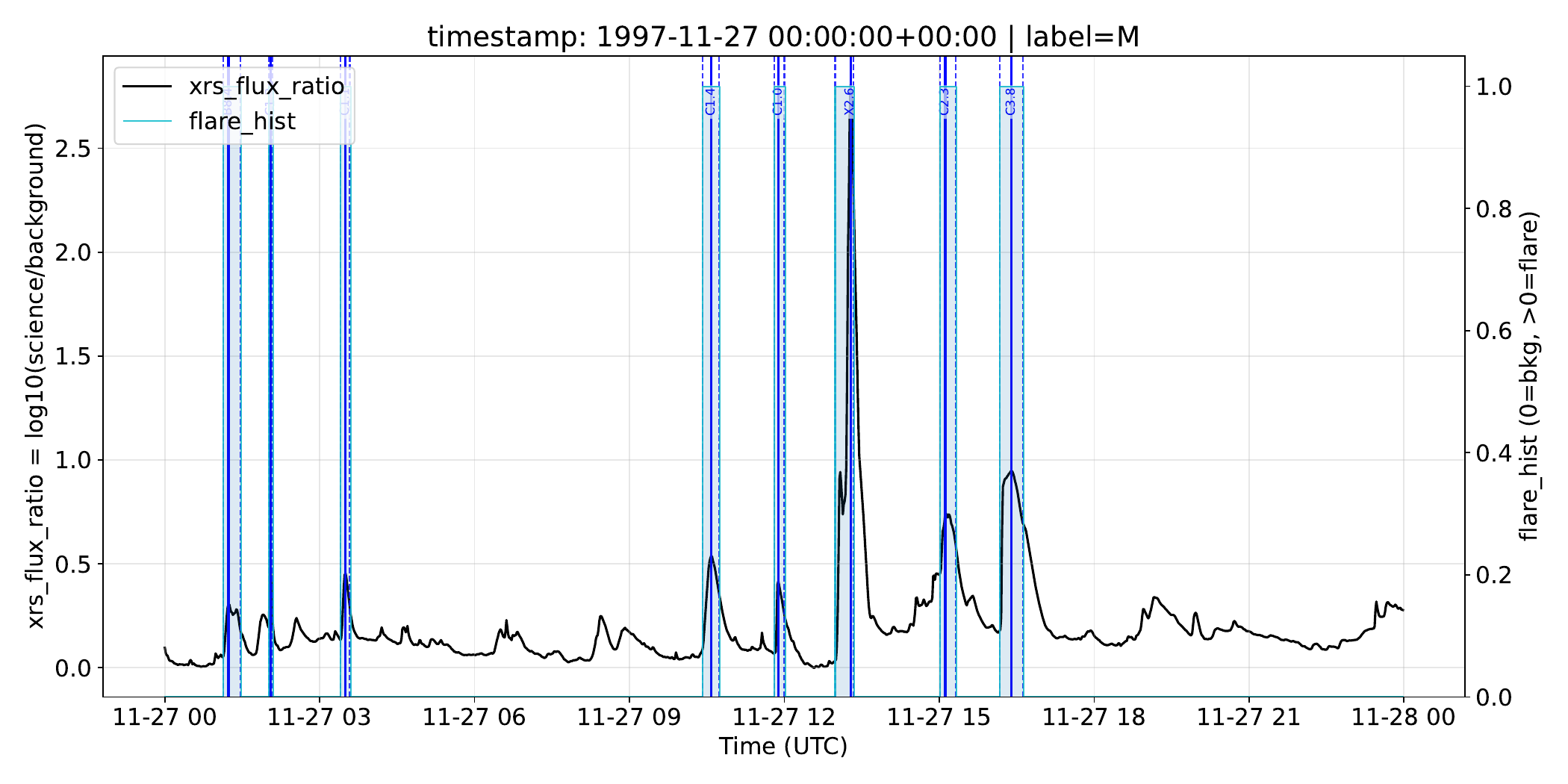}%
\caption{Sample XRS flux-ratio time series (black) spanning a 24-hour interval and labeled as M-class. Historical flare events occurring within the same interval are overplotted in blue.}%
\label{fig:xrs-sample}
\end{figure*}

\subsection{Video dataset}
\label{subsec:video-data}
The video dataset was constructed from the image data described above by grouping consecutive frames into sequences of $N_{\mathrm{frames}}$ images, each separated by a temporal interval of $\Delta t$ minutes. We adopted $N_{\mathrm{frames}}$ = 16, matching the standard input length used by most pretrained video models such as VideoMAE (see Section~\ref{subsec:model-videos}). Two temporal sampling cadences were considered, $\Delta t$ = 36 and 72 minutes, corresponding to total sequence durations $T$ of approximately 9.6 hours and 19.2 hours, respectively. Each video sample was assigned the label of its last frame, representing the flare class (or absence thereof) associated with the final timestamp in the sequence.

The number of video samples available for each flare class is reported in Table~\ref{tab:dataset}. Following the same strategy adopted for the image data, we also generated a downsampled training subset to mitigate class imbalance, which was subsequently used for model training.

\subsection{Time series dataset}
\label{subsec:timeseries-data}
The time series dataset was constructed from historical solar irradiance measurements collected by the GOES\footnote{\url{https://www.ncei.noaa.gov/products/goes-r-extreme-ultraviolet-xray-irradiance}}\footnote{\url{https://www.ncei.noaa.gov/products/goes-1-15/space-weather-instruments}}, covering nearly three decades of solar activity from 1995 onward. 
Specifically, we used the science-quality Level~2 (L2) series of 1-minute averaged soft X-ray fluxes in the long-wavelength band (0.1–0.8~nm; XRS-B channel). For each GOES satellite (08-18), the science fluxes were normalized by their corresponding daily background flux level (reported by GOES), producing a flux-ratio time series (science/background). These normalized series were then segmented into non overlapping 12h and 24h intervals (respectively containing 720 and 1440 entries with 1-minute cadence), each representing a distinct observation sample for forecasting. Samples with more than 70\% missing data were discarded, while shorter data gaps were filled by linear interpolation. Finally, a logarithmic transformation was applied to the flux ratios.

We enriched the time series dataset using historical flare event information from the NOAA SWPC Event Reports\footnote{\url{ftp://ftp.swpc.noaa.gov/pub/warehouse/}}, which provide start and end times, flare class, and associated active region identifiers (when available). For each XRS flux-ratio series, we generated a corresponding binary sequence with the same temporal cadence (1-minute), where each time step was assigned a value of 1 if a flare of any class occurred within that minute, and 0 otherwise. 

For each selected interval ($T$=12 h, 24 h), this process yielded two-channel time series inputs for the forecasting models: (i) the normalized XRS flux ratio (\textit{xrs-flux-ratio}), and (ii) the corresponding binary flare occurrence history (\textit{flux-hist}). The same historical flare catalog was also used to define the forecasting labels for each time series, following the same scheme adopted for the image and video datasets—indicating whether a C-, M-, or X-class flare occurred within the 24-hour period following the last recorded time step. An example of an \textit{xrs-flux-ratio} observation labeled as M-class is shown in Fig.~\ref{fig:xrs-sample}, where the corresponding historical flare events are overplotted in blue.

\begin{figure}[!htb]
\centering%
\includegraphics[scale=0.5]{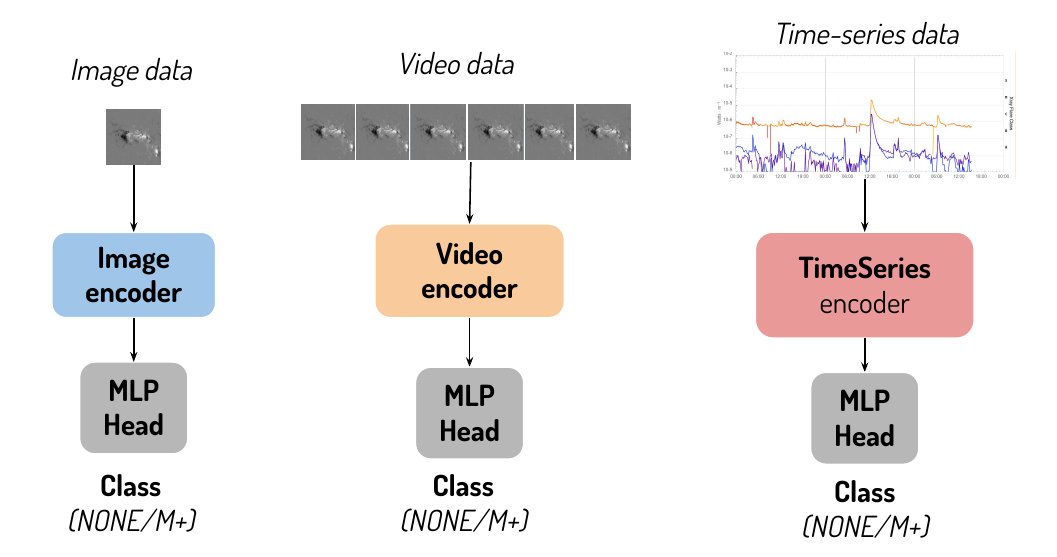}%
\caption{Overview of the three modality-specific solar flare forecasters. Each model processes a different data modality: (left) magnetogram images through a SigLIP2 vision transformer, (center) short sequences of magnetogram frames through a VideoMAE spatio-temporal transformer, and (right) soft X-ray flux time series from GOES through a Moirai2 temporal transformer. All models have the same classification head architecture for consistent comparison of forecasting skill across modalities.}%
\label{fig:model-schema}
\end{figure}

\section{Method}
\label{sec:method}

\subsection{Forecasting models}
\label{subsec:model}
Our forecasting framework is implemented using the \texttt{transformers}\footnote{\url{https://github.com/huggingface/transformers}} and \texttt{PyTorch}\footnote{\url{https://github.com/pytorch/pytorch}} libraries and supports three distinct data modalities: images, videos, and time series. As illustrated in Fig.~\ref{fig:model-schema}, each modality is processed by a dedicated backbone model designed to extract domain-relevant representations from its respective input type:
\begin{itemize}
\item \emph{Image-based model}: Individual solar magnetogram images are processed by a vision encoder (e.g., transformer-based architecture) that extracts spatial features describing the magnetic field morphology of active regions.
\item \emph{Video-based model}: Sequences of magnetogram frames are processed through a video backbone that captures both spatial and temporal dependencies, enabling the encoder to represent the dynamic evolution of active regions over time.
\item \emph{Time-series model}: Solar irradiance and related measurements (e.g., X-ray flux or flux ratios) are modeled using transformer-based sequence encoders that learn temporal dependencies and compress them into latent representations.
\end{itemize}
For all modalities, the resulting feature vector—encoding spatial, spatio-temporal, or purely temporal patterns—is passed to a fully connected classification head that predicts the probability of flare occurrence within a 24-hour forecasting window. The classification head consists of a linear layer producing a single output logit for binary forecasting tasks (e.g., \texttt{NONE} vs. \texttt{M+} or \texttt{NONE} vs. \texttt{C+}) or $K$ output logits for single- or multi-label $K$-class forecasting (e.g., \texttt{NONE}, \texttt{C}, \texttt{M}, \texttt{X}).

\subsubsection{Image model}
\label{subsec:model-images}
For the image-based forecasting task we employ \textit{SigLIP2} \citep{Tschannen2025siglip2}, a recent family of vision–language encoders that builds upon the original \textit{SigLIP} architecture \citep{Zhai2023siglip}. SigLIP builds on the \textit{Contrastive Language–Image Pretraining} (CLIP) model \citep{Radford2021}, replacing the original contrastive objective with a sigmoid loss, which improves alignment between visual and textual modalities during pretraining. This formulation improves training stability and yields stronger alignment between modalities. SigLIP2 further extends this approach by integrating captioning-based pretraining, self-distillation, masked prediction, and active data curation into a unified training recipe. These additions lead to more semantically rich and spatially precise visual representations, which are especially valuable for tasks such as solar flare forecasting.

The vision encoder of SigLIP2 follows the ViT architecture, partitioning an image into non-overlapping patches, linearly projecting them into embeddings, and modeling global dependencies through self-attention layers. In SigLIP2, training is carried out on the large-scale WebLI dataset \citep{WebLI} comprising approximately 10 billion images and 12 billion associated texts across 109 languages. This massive multimodal pretraining endows the encoder with robust and transferable visual features. Although the language encoder is not used in our experiments, the visual backbone benefits from this large-scale multimodal supervision.

To adapt SigLIP2 for flare classification, we discard the multimodal projection layers and attach a task-specific classification head on top of the pooled visual representation. The head consists of a linear mapping that outputs logits over the flare classes of interest. During fine-tuning, we experiment with both frozen and trainable vision backbones, enabling us to evaluate how much domain-specific specialization benefits solar forecasting. 

Practically, our implementation relies on the Hugging Face \texttt{AutoModelForImageClassification} interface, which allows us not only to load SigLIP2 checkpoints but also to seamlessly substitute alternative pretrained vision encoders that support the same API. This design ensures flexibility and extensibility for future comparisons across image backbones.

\begin{table*}[!ht]
\centering%
\setlength{\aboverulesep}{0pt}\setlength{\belowrulesep}{0pt}
\footnotesize%
\setlength{\extrarowheight}{2pt}
\caption{Summary of hyperparameter values used for training our forecaster models.}
\begin{tabular}{llll}
\hline%
\hline%
\rowcolor{pistachio}
\multirow{2}{*}{\cellcolor{pistachio}} & \multicolumn{3}{c}{Models}\\%
\arrayrulecolor{pistachio}\midrule\noalign{\vspace*{-\cmidrulewidth}}%
\arrayrulecolor{black}%
\cmidrule(lr){2-4}%
\rowcolor{pistachio}
\multirow{-2}{*}{Parameter} & SigLIP2 & VideoMAE & Moirai2\\%
\midrule
Input data size & \scriptsize{[3,224,224]} & \scriptsize{[16,3,224,224]} & \scriptsize{[1440,2]}\\%
Preprocessing & RandomCenterCrop(0.65,1.0) & RandomCenterCrop(0.65,1.0) & Log stretch xrs\_flux\_ratio\\%
& Resize(224$\times$224) & Resize(224$\times$224) & \\%
& RandomFlip & RandomFlip & \\%
& RandomRotate90 & RandomRotate90 & \\%
Model base & \scriptsize{\texttt{google/siglip2-base-patch16-224}} & \scriptsize{\texttt{MCG-NJU/videomae-base}} & \scriptsize{\texttt{Salesforce/moirai-2.0-R-small}} \\%
Frozen encoder layers & 10 & 10 & None \\%
\#GPUs & 4 & 4 & 1\\%
Epoch & 10 & 10 & 100\\%
Batch size & 64 & 16 & 64\\%
Gradient accumulation & 16 & 16 & 16\\%
Optimizer & AdamW & AdamW & AdamW\\%
Learning rate & 1e-4 & 1e-4 & 1e-4\\%
LR scheduler & cosine & cosine & cosine \\%
Warmup ratio & 0.3 & 0.3 & 0.1 \\%
\hline%
\hline%
\end{tabular}
\label{tab:model-pars}
\end{table*}

\subsubsection{Video model}
\label{subsec:model-videos}
For video-based forecasting we considered \textit{VideoMAE}\footnote{Video Masked Autoencoder, \url{https://github.com/MCG-NJU/VideoMAE}} model \citep{VideoMAE}, a transformer-based architecture specifically designed for spatio-temporal representation learning in videos. VideoMAE extends the masked autoencoding paradigm from images to videos by introducing \emph{tube masking}, where spatio-temporal cubes of a video are randomly masked at an extremely high ratio (90–95\%). The model is trained in a self-supervised manner to reconstruct the missing patches from the remaining visible tokens. This strategy forces the encoder to capture high-level structures and motion dynamics rather than relying on pixel-level redundancy, making it particularly suitable for solar magnetogram sequences in which the temporal evolution of active regions is essential.

The encoder of VideoMAE is based on a ViT architecture that jointly models space–time dependencies. Input video clips are divided into spatio-temporal tokens (cubes), which are embedded and processed through multi-head self-attention layers. This allows the model to simultaneously capture fine-grained spatial patterns (magnetic field morphology) and their temporal evolution across frames. During pretraining, only the encoder processes the visible tokens, while a lightweight decoder reconstructs the masked tokens; at fine-tuning, the decoder is discarded and the encoder serves as a general-purpose video backbone.

A crucial aspect of VideoMAE's success lies in its ability to exploit large-scale pretraining datasets. The original version demonstrated that competitive video representations could be obtained from relatively small datasets such as Kinetics-400 ($\sim$240k videos) \citep{Kai2017} or Something-Something V2 ($\sim$170k videos) \citep{Goyal2017}. VideoMAE v2 pushed this further by introducing large-scale pretraining on more than 1.3 million unlabeled video clips collected from diverse sources (YouTube, movies, web videos, and Instagram), complemented by a hybrid labeled dataset ($\sim$660k clips) for post-pretraining. This scaling was key to supporting billion-parameter video transformers and improving transferability across downstream tasks.

Recent improvements introduced in VideoMAE v2 \citep{VideoMAEv2} also make the approach more efficient and scalable. First, a \emph{dual masking strategy} was proposed, extending masking to the decoder in addition to the encoder. This reduces the number of tokens processed during reconstruction and nearly halves memory and computation costs while preserving accuracy. Second, VideoMAE v2 demonstrated that scaling to billion-parameter video transformers becomes feasible when coupled with large and diverse multi-source datasets (over one million clips), and a \emph{progressive training pipeline} that includes both unsupervised pretraining and supervised post-pretraining on hybrid datasets. Together, these innovations improve the generalization of the learned representations and allow the backbone to adapt more effectively to downstream tasks.

In our solar flare forecasting setup, we leverage the pretrained VideoMAE encoder and replace the reconstruction decoder with a classification head. Specifically, the pooled encoder representation of a magnetogram video is connected to a linear layer producing logits over the target flare classes. The model is then fine-tuned on labeled solar video sequences. This enables the backbone to adapt its spatio-temporal representations to the forecasting task, exploiting both morphological features of active regions and their dynamic evolution over time.

\subsubsection{Time series model}
\label{subsec:model-ts}
For time-series forecasting we employ \textit{Moirai}\footnote{Masked EncOder-based UnIveRsAl TIme Series Forecasting Transformer, \url{https://github.com/SalesforceAIResearch/uni2ts}}, a recently proposed foundation model for universal time-series forecasting \citep{woo2024moirai}. Moirai is built on a Transformer encoder architecture but introduces several innovations that make it particularly effective for heterogeneous, long, and irregularly sampled sequences. Its design combines (i) \textit{multi patch-size projections}, which enable efficient handling of both high- and low-frequency data, (ii) \textit{Any-variate Attention}, which flattens multivariate inputs and encodes both time and variate axes so that the model can flexibly process any number of input channels, and (iii) \textit{mixture distribution heads}, which allow probabilistic forecasting with flexible distributions beyond simple Gaussian assumptions. 

Moirai was pretrained on the \textit{Large-scale Open Time Series Archive} (LOTSA) \citep{woo2024moirai}, a corpus of over 27 billion observations across nine domains (energy, transport, climate, sales, finance, healthcare, etc.), making it one of the largest open pretraining efforts for time series. This large-scale pretraining enables the model to capture universal temporal patterns and transfer them to downstream tasks.

For our solar flare forecasting task, we adapt the pretrained Moirai encoder to process solar irradiance measurements (GOES XRS flux ratios and flare history). In our code, both the \textit{xrs\_flux\_ratio} and \textit{flare\_hist} are treated as \textit{variates} of the multivariate time series. We modified the \texttt{DataLoader} to handle our domain-specific JSON metadata format, and adjusted the \texttt{forward} method to support classification rather than probabilistic regression: instead of forecasting future values, the encoder outputs latent representations that are passed through a lightweight classification head (a linear projection into flare classes). 
The patching procedure follows Moirai’s standard approach: given an input of shape $[2,1440]$ (two variates, each with 1440 time steps), a patch size of 16 divides each variate into $1440/16=90$ patches, resulting in $2\times 90 = 180$ patch tokens. Each token corresponds to a contiguous window of 16 samples from a \emph{single} variate.
These tokens are embedded and processed jointly by the Moirai encoder, then pooled and passed to a lightweight classification head. Finally, we integrated these components into our training pipeline, adding various training losses and solar-specific evaluation metrics (see Section~\ref{subsec:model-training}).

By leveraging both its pretrained temporal representations from LOTSA and our tailored adaptations for solar physics data, Moirai provides a powerful backbone for flare forecasting, combining universal sequence modeling capabilities with domain-specific fine-tuning.

\subsection{Model training and evaluation}
\label{subsec:model-training}
In this work, we restricted our analysis to the prediction of solar flares with the greatest potential impact on Earth, focusing on the binary classification of \texttt{M+} flares (including both \texttt{M}- and \texttt{X}-class events) against the background class \texttt{NONE} (including observations with no flares or flares weaker than \texttt{M}-class). We trained multiple models, systematically varying the input data processing schemes and model hyperparameters to identify the optimal configuration for the forecasting task. The main hyperparameters adopted for all models are summarized in Table~\ref{tab:model-pars}.

All training experiments were conducted on the CINECA LEONARDO Booster supercomputing infrastructure\footnote{\url{https://www.hpc.cineca.it/systems/hardware/leonardo/}}, using a single compute node equipped with 4 GPUs (NVIDIA A100, 64 GB) and 1 CPU (Intel Xeon Platinum 8358 CPUs, 2.60 GHz), with 512 GB of RAM. The following subsection outlines the details of the adopted training strategy.

\paragraph{Input data and preprocessing}
The analysis was performed using the downsampled datasets described in Section~\ref{sec:dataset}, since preliminary experiments showed no significant difference in the prediction performance of the negative class when compared with the full original datasets. Image and video frame data were normalized to the range [0, 1]. For the time series data, a logarithmic transformation was applied only to the \textit{xrs\_flux\_ratio} channel, while the \textit{flare\_hist} channel was left unchanged.

\paragraph{Data augmentation} 
To enhance model generalization, we applied random data augmentation to the image and video datasets during training. Since the original dataset was constructed from full-disk images using a fixed cutout size (600$\times$600 pixels), rather than dynamically adapting the crop to the size of each AR, some solar structures unrelated to ARs or flare activity may be present. To mitigate this, we introduced a \texttt{RandomCenterCrop} transform that crops input images using a random crop fraction uniformly sampled in the range [0.65, 1.0]. The cropped images were then resized to match the input resolution required by the pretrained backbone (e.g., 224$\times$224 pixels). Additional augmentations included random horizontal and vertical flips, as well as random 90$^{\circ}$ rotations, each applied independently with equal probability. For validation and test data, only normalization and resizing transforms were applied to ensure consistency. The same augmentation pipeline was used for video data, with all transforms applied identically across all frames within each video sequence. No augmentation was applied to the time series data.

\paragraph{Training losses}
Because the solar flare datasets are highly imbalanced toward negative (non-flaring) examples, we implemented a custom Hugging Face (HF) trainer capable of rebalancing the training data within each batch.\footnote{When the number of negative samples greatly exceeds that of positive samples and the batch size is small, there is a high probability that a batch will be dominated $-$ or entirely composed $-$ of negative examples.} The trainer supports three alternative class-weighted loss functions to mitigate class imbalance: the class-weighted binary cross-entropy loss ($\mathcal{L}_{\mathrm{bce}}$), the class-weighted focal loss ($\mathcal{L}_{\mathrm{focal}}$), and the score-oriented loss (SOL; $\mathcal{L}_{\mathrm{sol}}$) introduced by \cite{Marchetti2022} and \cite{Guastavino2023}. A detailed mathematical description of these loss functions is provided in \ref{appendix:training-losses}.

\begin{table*}[!ht]
\centering%
\scriptsize%
\caption{Forecasting metrics for \texttt{M+} flare prediction obtained by fine-tuned SigLIP2 models on the HMI image validation set from \cite{Boucheron2023} (labeled as 'cv' in Table~\ref{tab:dataset}), using different training configurations and loss functions (binary cross-entropy, focal, and score-oriented TSS loss). Each metric is reported for two decision thresholds: $\tau=0.5$ and $\tau=\tau^{*}$, where $\tau^{*}$ is estimated from the full TSS–$\tau$ curves across multiple training and evaluation runs ($N=5$) as the threshold that maximizes the mean TSS, subject to minimum recall (0.5) and precision (0.15) constraints. For both thresholds, the table reports the mean and standard deviation of each metric computed over the $N=5$ runs.}
\begin{tabular}{cllclcccccc}
\hline%
\hline%
\multirow{2}[2]{*}{Model ID} & 
\multirow{2}[2]{*}{Train Set} &
\multirow{2}[2]{*}{Val Set} &
\multirow{2}[2]{*}{Sampler} &
\multirow{2}[2]{*}{Loss} &
\multicolumn{2}{c}{TSS} & \multicolumn{2}{c}{HSS} & \multicolumn{2}{c}{MCC}\\%
\cmidrule(lr){6-7}%
\cmidrule(lr){8-9}%
\cmidrule(lr){10-11}%
& & & & & $\tau=0.5$ & $\tau=\tau^{*}$ & $\tau=0.5$ & $\tau=\tau^{*}$ & $\tau=0.5$ & $\tau=\tau^{*}$\\%
\hline%
\texttt{siglip2\_v1} & train\_ds & cv & yes & sol & 0.680 $\pm$ 0.037 & 0.674 $\pm$ 0.037 & 0.211 $\pm$ 0.032 & 0.213 $\pm$ 0.031
 & 0.310 $\pm$ 0.024 & 0.310 $\pm$ 0.022\\%
\texttt{siglip2\_v2} & train\_ds & cv & no & focal\_w & 0.679 $\pm$ 0.022 & 0.656 $\pm$ 0.029  & 0.185 $\pm$ 0.021 & 0.213 $\pm$ 0.028 & 0.292 $\pm$ 0.018 & 0.306 $\pm$ 0.022\\%
\texttt{siglip2\_v3} & train\_ds & cv & no & bce\_w & 0.672 $\pm$ 0.019 & 0.620 $\pm$ 0.042 & 0.175 $\pm$ 0.031 & 0.210 $\pm$ 0.026 & 0.283 $\pm$ 0.020 & 0.295 $\pm$ 0.009\\%
\texttt{siglip2\_v4} & train\_ds\_bal & cv\_ds\_bal & no & sol & \textbf{0.695 $\pm$ 0.025} & 0.685 $\pm$ 0.025 & 0.207 $\pm$ 0.029 & 0.214 $\pm$ 0.031 & 0.311 $\pm$ 0.022 & 0.314 $\pm$ 0.024\\%
\texttt{siglip2\_v5} & train\_ds\_bal & cv\_ds\_bal & no & focal & 0.659 $\pm$ 0.055 & 0.640 $\pm$ 0.074 & 0.200 $\pm$ 0.037 & 0.211 $\pm$ 0.036 & 0.298 $\pm$ 0.023 & 0.301 $\pm$ 0.026\\%
\texttt{siglip2\_v6} & train\_ds\_bal & cv\_ds\_bal & no & bce & 0.682 $\pm$ 0.037 & \textbf{0.701 $\pm$ 0.028} & \textbf{0.239 $\pm$ 0.034} & \textbf{0.216 $\pm$ 0.036} & \textbf{0.330 $\pm$ 0.019} & \textbf{0.318 $\pm$ 0.021}\\%
\hline%
\hline%
\end{tabular}
\label{tab:image-metrics-cv}
\end{table*}

\begin{table*}[!ht]
\centering%
\scriptsize%
\caption{Forecasting metrics for \texttt{M+} flare prediction obtained by fine-tuned VideoMAE models on the HMI video validation set from \cite{Boucheron2023} (labeled as 'cv' in Table~\ref{tab:dataset}), using different training configurations. Each metric is reported for two decision thresholds: $\tau=0.5$ and $\tau=\tau^{*}$, where $\tau^{*}$ is estimated from the full TSS–$\tau$ curves across multiple training and evaluation runs ($N=5$) as the threshold that maximizes the mean TSS, subject to minimum recall (0.5) and precision (0.15) constraints. For both thresholds, the table reports the mean and standard deviation of each metric computed over the $N=5$ runs.}
\begin{tabular}{cclllcccccc}
\hline%
\hline%
\multirow{2}[2]{*}{Model ID} & 
\multirow{2}[2]{*}{$\Delta t$ (min)} & 
\multirow{2}[2]{*}{Train Set} &
\multirow{2}[2]{*}{Val Set} &
\multirow{2}[2]{*}{Loss} &
\multicolumn{2}{c}{TSS} & \multicolumn{2}{c}{HSS} & \multicolumn{2}{c}{MCC}\\%
\cmidrule(lr){6-7}%
\cmidrule(lr){8-9}%
\cmidrule(lr){10-11}%
& & & & & $\tau=0.5$ & $\tau=\tau^{*}$ & $\tau=0.5$ & $\tau=\tau^{*}$ & $\tau=0.5$ & $\tau=\tau^{*}$\\%
\hline%
\texttt{videomae\_v1} & 36 & train\_ds & cv\_ds & bce\_w & 0.678 $\pm$ 0.037 & \textbf{0.722 $\pm$ 0.038} & 0.281 $\pm$ 0.049 & 0.228 $\pm$ 0.050 & 0.360 $\pm$ 0.026 & 0.333 $\pm$ 0.041\\%
\texttt{videomae\_v2} & 36 & train\_ds\_bal & cv\_ds\_bal & bce & 0.696 $\pm$ 0.054 & 0.700 $\pm$ 0.044 & 0.266 $\pm$ 0.046 & 0.244 $\pm$ 0.047 & 0.354 $\pm$ 0.024 & 0.338 $\pm$ 0.026\\%
\texttt{videomae\_v3} & 72 & train\_ds & cv & bce\_w & 0.696 $\pm$ 0.044 & 0.710 $\pm$ 0.035 & 0.225 $\pm$ 0.072 & 0.211 $\pm$ 0.070 & 0.325  $\pm$ 0.050  & 0.319 $\pm$ 0.051\\%
\texttt{videomae\_v4} & 72 & train\_ds\_bal & cv\_ds\_bal & bce & \textbf{0.703 $\pm$ 0.036} & 0.711 $\pm$ 0.045 & \textbf{0.298 $\pm$ 0.034} & \textbf{0.273 $\pm$ 0.035} & \textbf{0.378 $\pm$ 0.017} & \textbf{0.362 $\pm$ 0.015}\\%
\hline%
\hline%
\end{tabular}
\label{tab:video-metrics-cv}
\end{table*}

\begin{table*}[!ht]
\centering%
\scriptsize%
\caption{Forecasting metrics for \texttt{M+} flare prediction obtained by fine-tuned Moirai2 models on the GOES XRS time series validation set (labeled as 'cv' in Table~\ref{tab:dataset}), using different training configurations. Each metric is reported for two decision thresholds: $\tau=0.5$ and $\tau=\tau^{*}$, where $\tau^{*}$ is estimated from the full TSS–$\tau$ curves across multiple training and evaluation runs ($N=5$) as the threshold that maximizes the mean TSS, subject to minimum recall (0.5) and precision (0.15) constraints. For both thresholds, the table reports the mean and standard deviation of each metric computed over the $N=5$ runs.}
\begin{tabular}{cclllcccccc}
\hline%
\hline%
\multirow{2}[2]{*}{Model ID} & 
\multirow{2}[2]{*}{$T$ (h)} & 
\multirow{2}[2]{*}{Train Set} &
\multirow{2}[2]{*}{Val Set} &
\multirow{2}[2]{*}{Loss} &
\multicolumn{2}{c}{TSS} & \multicolumn{2}{c}{HSS} & \multicolumn{2}{c}{MCC}\\%
\cmidrule(lr){6-7}%
\cmidrule(lr){8-9}%
\cmidrule(lr){10-11}%
& & & & & $\tau=0.5$ & $\tau=\tau^{*}$ & $\tau=0.5$ & $\tau=\tau^{*}$ & $\tau=0.5$ & $\tau=\tau^{*}$\\%
\hline%
\texttt{moirai2\_v1} & 12 & train & cv & bce\_w & 0.731 $\pm$ 0.020  & 0.736 $\pm$ 0.020 & 0.657 $\pm$ 0.079 & 0.635 $\pm$ 0.087 & 0.665 $\pm$ 0.071 & 0.648 $\pm$ 0.075\\%
\texttt{moirai2\_v2} & 24 & train & cv & bce\_w & \textbf{0.744 $\pm$ 0.027} & \textbf{0.747 $\pm$ 0.018} & \textbf{0.690 $\pm$ 0.026} & \textbf{0.680 $\pm$ 0.022} & \textbf{0.694 $\pm$ 0.026} & \textbf{0.686 $\pm$ 0.021}\\%
\hline%
\hline%
\end{tabular}
\label{tab:ts-metrics-cv}
\end{table*}

\paragraph{Model selection and evaluation}
For each sample $i=1,\dots,N$ with true target $y_i\in\{0,1\}$ (with $y_i=1$ denotes the occurrence of a flare), the model outputs a probability $\hat{p}_i\in[0,1]$\footnote{Model outputs for trained binary forecasters are single logits, converted to probabilities through a sigmoid activation function.} representing the likelihood of a flare occurring within the 24-hour forecasting window. The predicted class $\hat{y}_i$ is obtained by thresholding this probability at $\tau$ (=0.5 by default), i.e., $\hat{y}_i=\mathbbm{1}[\hat{p}_i\ge \tau]$. 
From the pairs $\{\hat{y}_i, y_i\}_{i=1}^N$ we computed the entries of the confusion matrix cin the table below: 

\begin{center}
\footnotesize%
\setlength{\tabcolsep}{4pt}
\begin{tabular}{ll|w{c}{0.6cm}|w{c}{0.6cm}|}
& \multicolumn{1}{c}{} & \multicolumn{2}{c}{\textbf{Predicted}} \\%
& \multicolumn{1}{c}{} & \multicolumn{1}{c}{\texttt{NONE}} & \multicolumn{1}{c}{\texttt{M+}}\\%
\cline{3-4}%
\multirow{2}{*}{\rotatebox[origin=c]{90}{\textbf{True}}} %
& \texttt{NONE} & $\mathrm{TN}$ & $\mathrm{FP}$ \\
\cline{3-4}%
& \texttt{M+} & $\mathrm{FN}$ & $\mathrm{TP}$ \\
\cline{3-4}%
\end{tabular}
\end{center}
where $\mathrm{TP}$, $\mathrm{TN}$, $\mathrm{FP}$, and $\mathrm{FN}$ denote the number of true positives, true negatives, false positives and false negatives, respectively. 

Binary forecasting performance was evaluated using the following metrics: \textit{True Skill Statistic} ($\mathrm{TSS}$), \textit{Heidke Skill Score} ($\mathrm{HSS}$), \textit{Matthews correlation coefficient} ($\mathrm{MCC}$).
Additional details and discussion of these metrics are provided in \ref{appendix:metrics}.

All metrics were computed on the validation dataset at the end of each training epoch, and the model achieving the highest TSS score was selected as the best-performing checkpoint. The TSS was chosen as the primary selection criterion because it is largely insensitive to class imbalance, in contrast to metrics such as the F1-score. The selected model was then evaluated on the held-out test set to derive the final forecasting metrics.

A comprehensive discussion of the advantages and limitations of different forecasting metrics is provided by \cite{Francisco2025}, who emphasize that HSS and MCC are more robust to variations in dataset composition. Accordingly, both metrics are reported alongside TSS when presenting the forecasting results in Section~\ref{sec:results}.

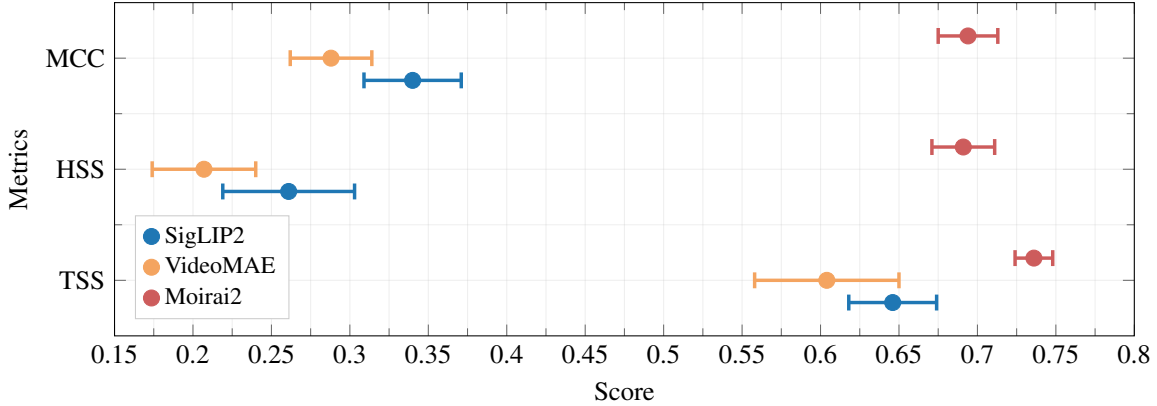
\begin{figure*}
\centering%
\pgfplotsset{
  compat=1.18,
  metricplot/.style={
    only marks,
    mark=*,
    mark size=3,
    line width=1pt,
    error bars/.cd,
      x dir=both,
      x explicit,
      error bar style={line width=1.3pt}, 
  }
}

\begin{tikzpicture}
\begin{axis}[
  width=15cm,
  height=6cm,
  xlabel={Score},
  ylabel={Metrics},
  xmin=0.15, xmax=0.8,
  ymin=0.5, ymax=3.5,
  ytick={1,2,3},
  yticklabels={TSS,HSS,MCC},
  grid=both,
  grid style={opacity=0.25},
  minor tick num=1,
  tick style={black!60},
  legend entries={SigLIP2,VideoMAE,Moirai2},
  legend style={
    at={(0.02,0.05)},
    anchor=south west,
    fill=white,
    draw=black!20,
    font=\small
  },
  legend cell align={left},
]

\addplot+[
  only marks, mark=*, mark size=3,
  draw=blue_mlp, mark options={draw=blue_mlp, fill=blue_mlp},
  /pgfplots/error bars/x dir=both,
  /pgfplots/error bars/x explicit,
  /pgfplots/error bars/error bar style={draw=blue_mlp, line width=1.3pt},
  /pgfplots/error bars/error mark options={draw=blue_mlp, line width=1.3pt},
  /pgfplots/error bars/error mark=|
]
coordinates {
  (0.646, 0.8) +- (0.028, 0)
  (0.261, 1.8) +- (0.042, 0)
  (0.340, 2.8) +- (0.031, 0)
};

\addplot+[
  only marks, mark=*, mark size=3,
  draw=orange_mlp, mark options={draw=orange_mlp, fill=orange_mlp},
  /pgfplots/error bars/x dir=both,
  /pgfplots/error bars/x explicit,
  /pgfplots/error bars/error bar style={draw=orange_mlp, line width=1.3pt},
  /pgfplots/error bars/error mark options={draw=orange_mlp, line width=1.3pt},
  /pgfplots/error bars/error mark=|
]
coordinates {
  (0.604, 1.0) +- (0.046, 0)
  (0.207, 2.0) +- (0.033, 0)
  (0.288, 3.0) +- (0.026, 0)
};

\addplot+[
  only marks, mark=*, mark size=3,
  draw=red_mlp, mark options={draw=red_mlp, fill=red_mlp},
  /pgfplots/error bars/x dir=both,
  /pgfplots/error bars/x explicit,
  /pgfplots/error bars/error bar style={draw=red_mlp, line width=1.3pt},
  /pgfplots/error bars/error mark options={draw=red_mlp, line width=1.3pt},
  /pgfplots/error bars/error mark=|
]
coordinates {
  (0.736, 1.2) +- (0.012, 0)
  (0.691, 2.2) +- (0.020, 0)
  (0.694, 3.2) +- (0.019, 0)
};
\end{axis}
\end{tikzpicture}
\caption{Forecasting metrics for \texttt{M+} flare prediction obtained with models trained on each data modality (blue: SigLIP2, orange: VideoMAE, red: Moirai2) and evaluated on the test set. For the image and video models, the box plot reports the average score and its standard deviation computed over $N=5$ independent runs, all trained and evaluated on the same original data split from \citet{Boucheron2023} (downsampled), labelled as "default" in Table~\ref{tab:metrics}. For the time series model, the same quantities are reported over $k=5$ independent data splits.}
\label{fig:metrics-test}
\end{figure*}

\section{Results}
\label{sec:results}
We began our analysis with the image dataset introduced by \cite{Boucheron2023}, aiming to identify optimal training parameters that could later be extended to the video models. As a first step, we conducted multiple training runs on the \textit{train\_ds} dataset (see Table~\ref{tab:dataset}) using different loss functions to determine the most effective configuration. Specifically, we evaluated a weighted binary cross-entropy loss, a weighted focal loss ($\gamma=2$), and a score-oriented loss based on the TSS metric. In the latter case, we employed a weighted random sampler to rebalance positive and negative samples within each batch, whereas for the other loss configurations, class imbalance was already compensated through loss weighting. Model selection across epochs was based on performance evaluated on the \textit{cv} dataset. We further investigated three additional configurations in which both training and model selection were performed on balanced downsampled datasets (\textit{train\_ds\_bal} and \textit{cv\_ds\_bal} in Table~\ref{tab:dataset}). In these cases, no class weighting or batch resampling was applied.

\begin{table*}[!ht]
\centering%
\scriptsize%
\caption{Forecasting metrics for \texttt{M+} flare prediction obtained with models trained on each data modality and evaluated on the test set. For the image and video models, the table reports the mean and standard deviation of the scores computed over $N=5$ independent runs, all trained and evaluated on the same original data split from \citet{Boucheron2023} (downsampled), labelled as "default". Additionally, scores averaged over $k=5$ alternative splits $-$ each randomly drawn from the original dataset as described in Section~\ref{sec:dataset} $-$ are reported and labelled as "alt.". For the time series model, metrics are likewise averaged over $k=5$ independent data splits.}
\begin{tabular}{llcccccc}
\hline%
\hline%
\multirow{2}[2]{*}{Model ID} & 
\multirow{2}[2]{*}{Data split} &
\multicolumn{2}{c}{TSS} & 
\multicolumn{2}{c}{HSS} & 
\multicolumn{2}{c}{MCC}\\%
\cmidrule(lr){3-4}%
\cmidrule(lr){5-6}%
\cmidrule(lr){7-8}%
& & $\tau=0.5$ & $\tau=\tau^{*}$ & $\tau=0.5$ & $\tau=\tau^{*}$ & $\tau=0.5$ & $\tau=\tau^{*}$\\%
\hline%
\multirow{2}{*}{\texttt{siglip2\_v6}} & default & \textbf{0.646 $\pm$ 0.028} & \textbf{0.671 $\pm$ 0.027} & \textbf{0.261 $\pm$ 0.042} & \textbf{0.235 $\pm$ 0.044} & \textbf{0.340 $\pm$ 0.031} & \textbf{0.329 $\pm$ 0.034} \\%
 & alt. & 0.645 $\pm$ 0.165 & 0.653 $\pm$ 0.161 & 0.197 $\pm$ 0.061 & 0.165 $\pm$ 0.074 & 0.278 $\pm$ 0.076 & 0.264 $\pm$ 0.079\\%
\hline%
\multirow{2}{*}{\texttt{videomae\_v4}} & default & \textbf{0.604 $\pm$ 0.046} & \textbf{0.613 $\pm$ 0.050} & \textbf{0.207 $\pm$ 0.033} & \textbf{0.191 $\pm$ 0.034} & \textbf{0.288 $\pm$ 0.026} & \textbf{0.278 $\pm$ 0.027}\\%
& alt. & 0.535 $\pm$ 0.073 & 0.569 $\pm$ 0.083 & 0.179 $\pm$ 0.084 & 0.167 $\pm$ 0.082 & 0.249 $\pm$ 0.058 & 0.249 $\pm$ 0.060 \\%
\hline%
\multirow{1}{*}{\texttt{moirai\_v2}} &  & 0.736 $\pm$ 0.012 & 0.740 $\pm$ 0.009 & 0.691 $\pm$ 0.020 & 0.683 $\pm$ 0.017 & 0.694 $\pm$ 0.019 & 0.688 $\pm$ 0.015\\%
\hline%
\hline%
\end{tabular}
\label{tab:metrics}
\end{table*}

The forecasting metrics (TSS, HSS, and MCC) obtained on the \textit{cv} dataset are summarized in Table~\ref{tab:image-metrics-cv}. For each model configuration (labeled as \texttt{siglip2\_v1-v6}), training was repeated $N=5$ times. Metrics were evaluated at two flare-decision thresholds: $\tau$=0.5 and $\tau=\tau^{*}$, where $\tau^{*}$ corresponds to the threshold that maximizes the mean TSS, estimated from the full TSS–$\tau$ curves across the $N$ runs under the constraint of a minimum recall of 0.5 and a minimum precision of 0.15. For both thresholds, we report the mean and standard deviation of each metric over the repeated runs. Within the run-to-run uncertainties, no statistically significant differences were observed among the different configurations. However, configuration \texttt{siglip2\_v6} $-$ trained on balanced datasets using the BCE loss $-$ achieved slightly higher values for most metrics and was therefore selected for the final evaluation of the image model on the test set.

We investigated two training configurations for the video models at each frame step size (36 and 72 minutes). The first used a weighted BCE loss on unbalanced training and validation datasets, while the second employed a standard BCE loss on balanced datasets. The forecasting metrics obtained on the validation sets for each configuration (labeled as \texttt{videomae\_v1-v4}) are summarized in Table~\ref{tab:video-metrics-cv} for different decision thresholds. Overall, model performances did not vary significantly across configurations. Slightly higher metric values were achieved with the 72-minute frame step and balanced datasets. Compared to the image-based models, the video models exhibited a modest improvement only in the HSS and MCC scores. Configuration \texttt{videomae\_v4} was therefore selected for the final evaluation on the video test set.

We trained the time series models on GOES XRS data using two alternative input sequence lengths, $T$ = 12 and 24 hours. The adopted model configurations, labeled as \texttt{moirai2\_v1-v2}, along with their corresponding validation metrics averaged over $N$=5 runs, are reported in Table~\ref{tab:ts-metrics-cv}. The model trained with the longer input window (\texttt{moirai2\_v2}), reaching slightly higher forecasting scores within the run-to-run uncertainties, was selected for the final evaluation.

Table~\ref{tab:metrics} summarizes the forecasting metrics obtained on the test set for the best-performing models selected through validation across all data modalities. For the image and video models, the table reports the mean and standard deviation of the scores computed over $N=5$ independent runs, each trained and evaluated on the same original downsampled data split from \citet{Boucheron2023}, labelled as "default". In addition, metrics averaged over $k=5$ alternative splits $-$ each randomly drawn from the original dataset as described in Section~\ref{sec:dataset} $-$ are also reported and labelled as "alt.". For the time series model, metrics are likewise averaged over $k=5$ independent data splits. 

The best performance metrics for each model are shown in Fig.~\ref{fig:metrics-test} as colored markers $-$ blue for SigLIP2, orange for VideoMAE, and red for Moirai2 $-$ indicating the mean scores with their corresponding standard deviations. The SigLIP2 image model achieved TSS=0.646, HSS=0.261, and MCC=0.340, confirming the ability of vision–language transformers to extract meaningful magnetic features from single magnetograms. The VideoMAE model, trained on 16-frame magnetogram sequences, yielded slightly lower scores (TSS=0.604, HSS=0.207, MCC=0.288), suggesting a limited ability to fully exploit temporal information and capture pre-flare evolution, even when varying cadence intervals (32, 72). 

Overall, both the image and video models achieve slightly higher metric values when evaluated on the original data split compared to the alternative samples. The Moirai2 time-series model achieved the best overall performance, with TSS=0.736, HSS=0.691, and MCC=0.694, outperforming both the image and video models, particularly for the HSS and MCC metrics.

The TSS scores obtained with single-image data ($\sim$0.65) are consistent with values reported in previous studies using SDO/HMI AR magnetograms for the same forecasting task (\texttt{M+} class prediction with a 24-hour horizon), despite differences in data splits and model architectures. In contrast, models trained on multi-wavelength full-disk data (e.g., \citealt{Sun2023, Francisco2025}) have achieved higher HSS and MCC values, typically around 0.4. For the video models, all tested configurations tend to slightly underperform on the test set compared with both our image-based models and recently published video-based approaches (e.g., \citealt{Li2025, Xu2025}). While the inclusion of temporal information is physically expected to provide a measurable advantage over single-image models, such improvements have generally been modest. Other studies (e.g., \citealt{Francisco2024}) report performance gains of $\sim$0.02–0.03 in skill scores $-$ often comparable to the typical run-to-run uncertainty. We therefore performed additional training runs to further investigate these results.

We first examined whether alternative video cadences and sequence lengths could yield performance improvements. In addition to the previously tested shorter sequences (frame cadence $\Delta t = 32$~min, total duration $T \approx 9.6$~h), we considered much longer sequences ($\Delta t = 288$~min, $T \approx 3.2$~days). The latter configuration reduced the size of the training dataset by a factor of 2.6 compared to the 72-minute cadence case. Models were trained and evaluated under the same conditions as the selected \texttt{videomae\_v4} configuration. In both cases, the resulting HSS and MCC metrics on the test set were comparable, while the TSS scores were lower: 0.565~$\pm$~0.059 for $\Delta t = 36$~min and 0.507~$\pm$~0.062 for $\Delta t = 288$~min.

Since no significant differences were observed across alternative video data settings, we next investigated whether performance could be improved on the $\Delta t$=72 min dataset through alternative hyperparameter configurations. One possible explanation for the limited gains is that the pretrained VideoMAE representations $-$ learned from approximately 300,000 natural videos $-$ may transfer less effectively to the solar domain compared to SigLIP2, which was pretrained on a much larger dataset of about 10 billion images. Because a large portion of the encoder weights was frozen during fine-tuning, this may have constrained the model's capacity to adapt to the solar data. To test this hypothesis, we performed a new experiment in which the entire VideoMAE encoder was fine-tuned. To further mitigate overfitting, we introduced a dropout layer ($p=0.3$) in the classification head and applied weight decay ($\lambda=0.07$) relative to the previous configurations. The resulting model reached comparable HSS and MCC scores and a slightly better TSS (0.636 $\pm$ 0.046), still compatible with previous results within the run-to-run uncertainties.

We also tested alternative learning rate and warm-up settings, reducing the learning rate from 1e-4 to 5e-5 and varying the warm-up ratio between 0.1 and 0.3. These configurations, however, yielded forecasting scores comparable to the default setup. 


\section{Summary}
\label{sec:summary}
In this work, we conducted a systematic evaluation of recent transformer-based architectures applied to solar flare forecasting across three complementary data modalities $-$ images, videos, and time series. Using publicly available datasets of SDO/HMI magnetograms and GOES soft X-ray fluxes, we benchmarked the \textit{SigLIP2}, \textit{VideoMAE}, and \textit{Moirai2} models, each representing a state-of-the-art transformer design in its respective domain. All models were trained and validated under consistent data splits, balanced sampling strategies, and identical evaluation protocols based on the TSS, HSS, and MCC forecasting metrics. All trained model weights, software code, and datasets have been publicly released to promote reproducibility and collaboration, supporting the development of next-generation, physics-informed, and operationally deployable flare forecasting systems. 

From a physical standpoint, the comparative behavior of the three transformer models reflects the distinct nature of the information encoded in each data modality. The SigLIP2 model, trained on single-frame magnetograms, captures the instantaneous spatial distribution of magnetic fields in active regions $-$ primarily the polarity inversion lines, magnetic shear, and $\delta$-spot structures $-$ which are well-established indicators of stored free magnetic energy and potential flare productivity \citep[e.g.,][]{Rom14, Rom18, Rom24}. However, without temporal context, the model remains sensitive only to static morphological proxies of magnetic non-potentiality.

The VideoMAE architecture extends this representation to the temporal domain, tracking the short-term evolution of active regions and thus probing the processes of flux emergence, shear accumulation, and magnetic cancellation that precede energy release. Nevertheless, its forecasting gain over static images remains moderate. This suggests that, within the $\sim$10–20 h windows adopted, the temporal variability of the photospheric field still traces quasi-stationary phases of magnetic energy build-up, while the rapid reconfiguration associated with reconnection occurs on shorter coronal timescales not fully sampled by the input sequences. For longer temporal windows ($\sim$3~days), the limited size and cadence of the available video dataset likely prevent the model's ability to learn these subtle pre-eruptive dynamics.

The superior performance of the Moirai2 time-series model highlights instead the predictive value of irradiance-based temporal information. The GOES X-ray flux encodes the integrated coronal response to energy release and heating, thus serving as a direct tracer of magnetic energy dissipation. The model’s ability to identify flare precursors from smooth irradiance variations and flare history suggests that the temporal evolution of coronal emission already contains the statistical imprint of the magnetic processes governing flare occurrence. In this sense, Moirai2 acts as a data-driven proxy for the coronal energy storage and release cycle, whereas image and video models probe only its boundary conditions at the photosphere.

Overall, the comparative results indicate that magnetic morphology alone provides a necessary but not sufficient condition for flare occurrence, while irradiance time series effectively encode the integrated temporal signature of the coronal response. A future unified multimodal architecture combining both the spatial topology of magnetic fields and the temporal evolution of coronal emission could thus enable physically consistent, end-to-end forecasting of solar eruptive activity.


Building on these findings, we plan to extend this work to a revised and enriched multimodal dataset integrating full-disk video sequences of HMI and EUV observations across multiple wavelengths, together with multivariate time series from GOES irradiance sensors and other observables (e.g., AR features). The new dataset will include additional metadata and diagnostic flags $-$ currently unavailable in the \citet{Boucheron2023} dataset $-$ enabling more comprehensive investigations, such as distinguishing the influence of single versus multiple active regions and assessing event directionality and potential geoeffectiveness at Earth.

Future developments will also explore generative modeling (e.g., GANs or diffusion-based models) to expand and rebalance the training dataset, and self-supervised pretraining of video models to enhance temporal representation learning. These efforts aim to overcome current data limitations and move toward a robust, physics-aware multimodal architecture for operational solar flare forecasting.

\section*{Acknowledgements}
\label{sec:acknowledgements}
\small{
Supported by Italian Research Center on High Performance Computing Big Data and Quantum Computing (ICSC), project funded by European Union - NextGenerationEU - and National Recovery and Resilience Plan (NRRP) - Mission 4 Component 2 within the activities of Spoke 3 (Astrophysics and Cosmos Observations).

We acknowledge ISCRA for awarding this project access to the LEONARDO supercomputer, owned by the EuroHPC Joint Undertaking, hosted by CINECA (Italy). Additional computing resources for this work were provided by the INAF "CIRASA" (\emph{Collaborative and Integrated platform for Radio Astronomical Source Analysis}) project, and the Italian PNRR Project IR0000034 "STILES" (\emph{Strengthening the Italian leadership in ELT and SKA}) project. S.R. acknoledges the INAF SCIARADA grant for financial support.
}%

\section*{Data Availability}
\label{sec:data-availability}
\small{
The software code used in this work is publicly available under the GNU General Public License v3.0\footnote{\scriptsize{\url{https://www.gnu.org/licenses/gpl-3.0.html}}} on the GitHub repository \url{https://github.com/inaf-oact-ai/solar-flare-forecaster/}. 

The datasets and trained model weights have been made available in this \emph{Hugging Face} repository: \url{https://huggingface.co/inaf-oact-ai}.
}







\newpage%

\appendix%
\setcounter{figure}{0}
\setcounter{table}{0}

\section{Training losses}
\label{appendix:training-losses}
For single-label multi-class forecasting, the framework currently supports these possible losses:
\begin{itemize}
\item \emph{Cross entropy (CE) loss}: $\mathcal{L}_{\text{ce}}$ 
\begin{equation*}
\mathcal{L}_{\text{ce}}= \frac{1}{N} \sum_{i=1}^{N} \sum_{k=0}^{K-1} w_k \, \mathbbm{1}\{ y_{i} = k \} \, \Biggl( - \log \frac{\exp(z_{ik})}{\sum_{j=0}^{K-1} \exp(z_{ij})} \Biggr).
\end{equation*}
\item \emph{Focal loss}: $\mathcal{L}_{\text{focal}}$
\begin{equation*}
\mathcal{L}_{\text{focal}}= \frac{1}{N} \sum_{i=1}^{N} \sum_{k=0}^{K-1} 
w_k \, \mathbbm{1}\{y_{i} = k\} \, 
\bigl(1 - p_{ik} \bigr)^{\gamma} \, \log(p_{ik}).
\end{equation*}
\item \emph{Score-oriented loss}: $\mathcal{L}_{\text{sol}}$
\begin{eqnarray*}
\mathcal{L}_{\text{sol}}&=& -\frac{1}{K}\sum_{k=0}^{K-1}\mathrm{TSS}_k\\
\mathrm{TSS}_k &=& \frac{\mathrm{TP}_k}{\mathrm{TP}_k+\mathrm{FN}_k} + \frac{\mathrm{TN}_k}{\mathrm{TN}_k+\mathrm{FP}_k} - 1\\
\mathrm{TP}_k &=& \sum_{i=1}^N y_{ik}\,p_{ik}\\
\mathrm{TN}_k &=& \sum_{i=1}^N (1-y_{ik})\bigl(1-p_{ik}\bigr)\\
\mathrm{FP}_k &=& \sum_{i=1}^N (1-y_{ik})\,p_{ik}\\
\mathrm{FN}_k &=& \sum_{i=1}^N y_{ik}\bigl(1-p_{ik}\bigr)
\end{eqnarray*}
\end{itemize}
where $N$ is the batch size (i.e. number of training samples in the batch), $K$ is the number of classes, $z_{ik}$ are the model logits for sample $i$ and class $k$, $y_i\in\{0,\dots,K-1\}$ are the true class indices with $y_{ik}$ their one-hot encoding, $p$ is the predicted probability of the true class, $w_k$ are the class weights, $\gamma$ is a focusing parameter, and $\mathbbm{1}\{\cdot\}$ is the indicator function. $\mathrm{TSS}$ denote the \textit{True Skill Statistic} ($\in$[-1,1]), while $\mathrm{TP}$, $\mathrm{TN}$, $\mathrm{FP}$, and $\mathrm{FN}$ are the number of true positives, true negatives, false positives and false negatives, respectively. Similarly, the framework provides the corresponding losses for the binary forecasting task.\\ 
The class weights $w_k$ are pre-computed according to the forecasting task:
\begin{align*}
\text{(Multiclass)} \quad & w_k = \frac{N}{K \, n_k}, \\[6pt]
\text{(Binary)} \quad & w_k = \frac{n_{neg}}{n_{pos}}
\end{align*}
where $N$ is the total number of samples, $n_k$ the number of samples belonging to class $k$, $n_{neg}$ and $n_{pos}$ are the number of negatives and positive samples, respectively.

\section{Forecasting metrics}
\label{appendix:metrics}
\paragraph{F1-score}
The F1 score summarizes the balance between misses and false alarms at a fixed threshold by the harmonic mean of precision $\mathcal{P}$ and recall $\mathcal{R}$:
\begin{eqnarray}
\mathrm{F1}&=&2\,\frac{\mathcal{P}\times\mathcal{R}}{\mathcal{P}+\mathcal{R}}=\frac{2\,\mathrm{TP}}{2\,\mathrm{TP}+\mathrm{FP}+\mathrm{FN}}
\end{eqnarray}
where the recall $\mathcal{R}$ (or true positive rate/sensitivity $\mathrm{TPR}$) and the precision $\mathcal{P}$ are respectively computed as:
\begin{eqnarray}
\mathcal{R}&=&\mathrm{TPR}=\frac{\mathrm{TP}}{\mathrm{TP}+\mathrm{FN}}\\
\mathcal{P}&=&\frac{\mathrm{TP}}{\mathrm{TP}+\mathrm{FP}}%
\end{eqnarray}
F1 ranges in $[0,1]$, with $1$ indicating perfect classification.

\paragraph{True Skill Statistic (TSS)} 
The TSS score, also known as \textit{Hanssen \& Kuipers’ Discriminant} (H\&KSS) \citep{Hanssen1965} measures the forecast's ability to correctly predict both positive (event) and negative (non-event) outcomes:
\begin{equation}
\mathrm{TSS} \;=\; \frac{\mathrm{TP}}{\mathrm{TP}+\mathrm{FN}} \;-\; \frac{\mathrm{FP}}{\mathrm{FP}+\mathrm{TN}}
\;=\; \mathrm{TPR}+\mathrm{TNR}-1
\end{equation}
where $\mathrm{TNR}$ is the true negative rate (specificity), computed as:
\begin{eqnarray}
\mathrm{TNR}&=&\frac{\mathrm{FP}}{\mathrm{FP}+\mathrm{TN}}
\end{eqnarray}
TSS ranges from $-1$ (perfectly wrong) through $0$ (no skill, random/constant forecast) to $1$ (perfect). TSS is widely used in space-weather because it is relatively insensitive to class imbalance, while other metrics like the F1-score are more affected when the proportion of the minority class (rare event) changes.

\paragraph{Heidke Skill Score (HSS)} 
HSS score, like TSS, includes all confusion matrix entries to measure the
forecasting accuracy relative to random chance:
\begin{equation}
\mathrm{HSS} \;=\;
\frac{2\;\big(\mathrm{TP}\times\mathrm{TN}-\mathrm{FN}\times\mathrm{FP}\big)}
{(\mathrm{TP}+\mathrm{FN})\times(\mathrm{FN}+\mathrm{TN})+(\mathrm{TP}+\mathrm{FP})\times(\mathrm{FP}+\mathrm{TN})}.
\end{equation}
Its range is $(-\infty,1]$, with $1$ perfect forecasting and $0$ indicating no improvement over chance. 
Unlike TSS, HSS is sensitive to a varying event/no-event sample ratio \citep{Bloomfield2012}. 

\paragraph{Matthews correlation coefficient (MCC)} 
MCC is the Pearson correlation between predicted and observed binary outcomes:
\begin{equation}
\mathrm{MCC} \;=\;
\frac{\mathrm{TP}\times\mathrm{TN}-\mathrm{FP}\times\mathrm{FN}}
{\sqrt{(\mathrm{TP}+\mathrm{FP})\times(\mathrm{TP}+\mathrm{FN})\times(\mathrm{TN}+\mathrm{FP})\times(\mathrm{TN}+\mathrm{FN})}}.
\end{equation}
It ranges from $-1$ (total disagreement) through $0$ (average random prediction) to $1$ (perfect prediction). 
MCC uses all four entries of the confusion matrix and is robust to imbalance, making it suitable when both classes are important.

\end{document}